# High Electron Mobility, Quantum Hall Effect and Anomalous Optical Response in Atomically Thin InSe

D. A. Bandurin[1], A. V. Tyurnina[2,3], G. L. Yu[1], A. Mishchenko[1], V. Zólyomi[2], S. V. Morozov[4,5], R. Krishna Kumar[2], R. V. Gorbachev[2], Z. R. Kudrynskyi[6], S. Pezzini[7], Z. D. Kovalyuk[8], U. Zeitler[7], K. S. Novoselov[2], A. Patanè[6], L. Eaves[6], I. V. Grigorieva[1], V. I. Fal'ko[1,2], A. K. Geim[1], Y. Cao[1,2]

[1]School of Physics, University of Manchester, Oxford Road, Manchester M13 9PL, United Kingdom
[2]National Graphene Institute, University of Manchester, Manchester M13 9PL, United Kingdom
[3]Skolkovo Institute of Science and Technology, Nobel St 3, 143026 Moscow, Russia
[4]Institute of Microelectronics Technology and High Purity Materials, RAS, Chernogolovka 142432, Russia
[5]National University of Science and Technology 'MISiS', Leninsky Pr. 4, 119049 Moscow, Russia
[6]School of Physics and Astronomy, University of Nottingham NG7 2RD, United Kingdom
[7]High Field Magnet Laboratory, Radboud University, Toernooiveld 7, 6525 ED Nijmegen, the Netherlands
[8]National Academy of Sciences of Ukraine, Institute for Problems of Materials Science, Chernovtsy, Ukraine

*A decade of intense research on two-dimensional (2D) atomic crystals has revealed that their properties can differ greatly from those of the parent compound[1,2]. These differences are governed by changes in the band structure due to quantum confinement and are most profound if the underlying lattice symmetry changes[3,4]. Here we report a high-quality 2D electron gas in few-layer InSe encapsulated in hexagonal boron nitride under an inert atmosphere. Carrier mobilities are found to exceed $10^3$ and $10^4$ $cm^2/Vs$ at room and liquid-helium temperatures, respectively, allowing the observation of the fully-developed quantum Hall effect. The conduction electrons occupy a single 2D sub-band and have a small effective mass. Photoluminescence spectroscopy reveals that the bandgap increases by more than 0.5 eV with decreasing the thickness from bulk to bilayer InSe. The band-edge optical response vanishes in monolayer InSe, which is attributed to monolayer's mirror-plane symmetry. Encapsulated 2D InSe expands the family of graphene-like semiconductors and, in terms of quality, is competitive with atomically-thin dichalcogenides[5,6,7] and black phosphorus[8,9,10,11].*

Indium selenide belongs to the family of layered metal-chalcogenide semiconductors. Each of its layers has a honeycomb lattice that effectively consists of 4 covalently bonded Se-In-In-Se atomic planes (see Fig. 1a). The layers are held together by van der Waals interactions at an interlayer distance $d \approx 0.8$ nm. The specific stacking order in bulk γ-InSe, where indium atoms in one layer are aligned with selenium atoms in the other, breaks down the mirror-plane symmetry characteristic for monolayer InSe. Earlier studies of bulk InSe revealed a small effective mass in the conduction band[12,13], a high electron mobility at room temperature (RT) due to weak electron-phonon scattering[14] and optical activity in absorption and emission[15]. More recently, there have been several reports on thin InSe films made by mechanical exfoliation. Optical studies[16,17,18] have proved that their band gap greatly varied with decreasing the number of layers N, in agreement with density functional theory (DFT) [18,19]. Few-layer InSe also exhibits promising characteristics for optoelectronic applications[16,20,21,22]. Furthermore, the 2D electron gas (2DEG) induced by the field effect at the surface of multilayer (≥20 nm) InSe crystals showed low-temperature mobilities μ up to 2,000 $cm^2/Vs$[23,24], approaching values typically found for the 2D accumulation layers formed near stacking faults in bulk InSe[14,25]. However, the scarcity of experimental data and spread in the reported characteristics suggest that atomically thin films suffer from considerable degradation with respect to bulk InSe, possibly due to reaction with chemical species present in air such as oxygen and water (Supplementary Section 6). To circumvent the problem of limited chemical stability, in this work we employ exfoliation and subsequent encapsulation[26,27] of few-layer InSe in an inert (argon) atmosphere[28]. This has allowed us to fabricate InSe structures and field-effect devices (FED) down to a monolayer, and they exhibited previously unattainable quality and stability under ambient conditions.

Figure 1b shows an optical image of a typical InSe flake mechanically exfoliated and visualized inside a glove box filled with argon[28]. The layer thickness was determined from the optical contrast and verified by atomic force microscopy (AFM). To fabricate our FED the exfoliated InSe crystal was transferred onto a hexagonal boron nitride (hBN) flake prepared on a SiO$_2$/doped-Si wafer. A few-layer ($N$ = 6–10 ) graphene (FLG) was positioned on top of the hBN to serve as electrical contacts to the InSe crystal[5], see Fig. 1c. All the structures were then covered with a second hBN crystal to completely isolate InSe from the environment. Mesas were etched through the top hBN, InSe and graphene to define Hall bars using the gold top gate as the etching mask (Fig. 1d). Finally, Au/Cr contacts to graphene were deposited as shown in Figs. 1c,d and further discussed in Methods. The top hBN served as the dielectric layer for the top gate whereas the SiO$_2$/Si wafer acted as the bottom dielectric/gate electrode (Fig. 1c). We studied the electrical properties of six multiterminal InSe devices with thicknesses ranging from 1 to 10 layers.

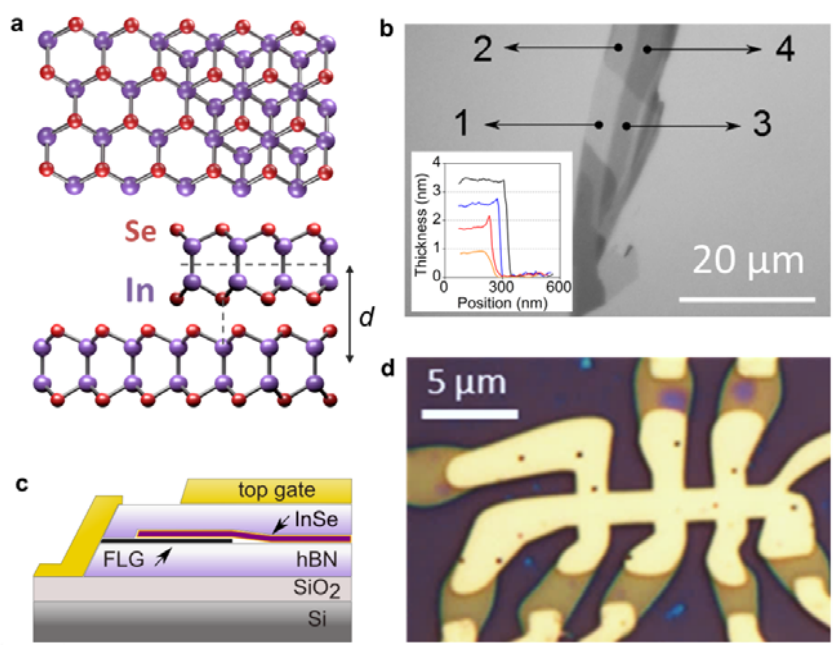

**Figure 1| 2D InSe devices. a,** Schematic representation of the monolayer and bilayer crystal structures. Purple and red spheres correspond to indium and selenium atoms, respectively. **b,** Micrograph of an InSe flake exfoliated onto a 300 nm thick polymer film. Numbers from 1 to 4 correspond to the number of layers, determined by AFM. Inset: Their AFM profiles with respect to the polymer substrate. **c,** Cross-sectional schematic of our FEDs. **d,** Optical micrograph of one of them. The central bright area is the top gate that covers encapsulated InSe. Peripheral Au contacts lead to hBN-encapsulated FLG not covered by the top gate; the FLG region shows in dark yellow.

By applying top and bottom gate voltages ($V_{tg}$ and $V_{bg}$, respectively) we could control the electron density $n$ in InSe over a wide range up to $\approx 10^{13}$ cm$^{-2}$. We found that the contact resistance between 2D InSe and FLG also depended on gate voltage (Supplementary Section 1). This can be attributed to changes in the Schottky barrier height that depends on doping. Reasonably good ohmic contacts (5–30 k$\Omega$ μm) were achieved for $n > 10^{12}$ cm$^{-2}$, allowing four-probe measurements using the standard lock-in techniques. In the four-terminal geometry, the sheet resistivity $\rho_{xx}$ of our few-layer InSe devices could be changed between approximately 100 Ω and few kΩ by varying $V_{bg}$ and $V_{tg}$. Much higher resistivities were achieved by applying negative gate voltages (removing electrons from the channel) but this allowed only two-probe measurements because of very high channel and contact resistances (Supplementary Section 3).

Examples of the dependence of $\rho_{xx}$ on $V_{bg}$ are shown in Fig. 2a for a 6-layer (6L) device. The resistivity increases with temperature $T$ for all the gate voltages, indicating metallic behavior. Carrier density $n$ was determined from Hall measurements, and its values for different $V_{bg}$ agree with the concentrations estimated from the device geometrical capacitance (inset of Fig. 2a). Fig. 2b shows the values of Hall mobility, $\mu = 1/ne\rho_{xx}$, as a function of $T$ for different $n$ accessible in our four-probe measurements, where $e$ is the electron charge. Below 50 K, µ is almost independent of $T$, being limited by disorder. Its screening by a high density of electrons leads to an increase in µ, and our best device exhibited an electron mobility of 12,700 cm²/Vs for $n \approx 8 \times 10^{12}$ cm⁻². For $T > 100$ K, we observed a gradual decrease in µ, somewhat faster than the standard $T^{-1}$ dependence expected for acoustic phonon scattering in the Bloch-Gruneisen regime[29]. This can be attributed to additional scattering at homopolar optical phonons with a low activation energy ≈13 meV (see Ref. [19]). At RT, µ drops to ≈1,000 cm²/Vs, which is higher than the highest mobility reported for 2D dichalcogenides[5,7] and comparable to that in black phosphorus[11]. Similar behavior was found for 3L and 10L devices (inset of Fig. 2b and Supplementary Section 4).

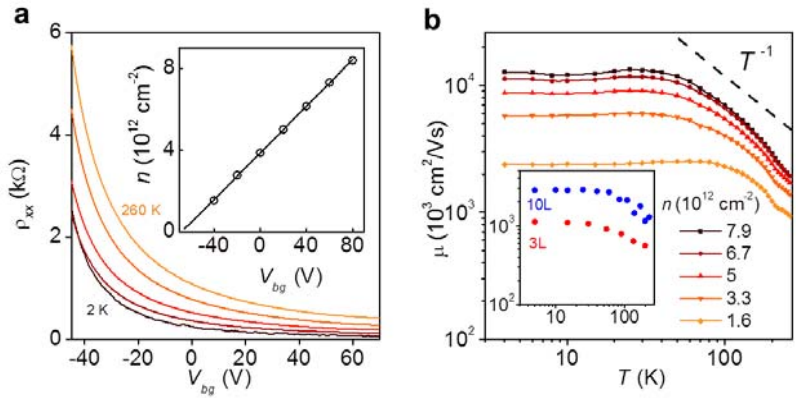

**Figure 2| Transport properties of atomically thin InSe. a,** The resistivity of a 6L device as a function of $V_b$ at different $T$. Additional top gate voltage ($V_{tg} = 8$ V) was applied to increase $n$ in the 2DEG. For negative gate voltages ($V_{bg} < -45$ V in this figure), the contact resistance increased above 100 kΩ, which did not allow four-probe measurements. Large on/off ratios, similar to those in the previous reports[23,24], could be measured in the 2-probe geometry (Supplementary Section 3). Inset: $n$ found from Hall measurements (symbols); the solid line shows the density expected from the known geometrical capacitance. **b,** $T$-dependence of the Hall mobility µ for the 6L device. The dashed black line shows the dependence expected for acoustic phonons. Inset: µ ($T$) for 3L and 10L at high doping (8.3 and $7.7 \times 10^{12}$ cm⁻², respectively).

Fig. 3 shows behavior of $\rho_{xx}$ and Hall resistivity $\rho_{xy}$ in magnetic field $B$ for our 3L and 6L devices (data for a 10L device are provided in Supplementary Section 4). In the 6L device, Shubnikov de Haas oscillations (SdHO) start at $B \approx 4.5$ T, which yields a quantum mobility of about 2,200 cm²/Vs, close to the Hall mobility measured for the same $n$ (Fig. 2b). In high $B$, SdHO developed into the quantum Hall effect (QHE) so that $\rho_{xx}$ diminished and $\rho_{xy}$ exhibited plateaus (Figs. 3a,f). Figures 3b,c provide further details about how the amplitude and phase of the SdHO evolve with $B$ and $n$. In particular, the Landau fan diagram in Fig. 3c plots the $B$ values in which minima of the oscillations occur. Each set of the minima can be extrapolated to zero (with an experimental uncertainty of ±15%), proving that SdHO in 2D InSe have the standard phase. This means that, unlike graphene, 2D InSe exhibits no Berry phase[3], in agreement with general expectations. Furthermore, the period of SdHO, $\Delta(1/B)$, allowed independent measurement of the electron density as $n = \frac{2e}{h}\frac{1}{\Delta(1/B)}$ where $h$ is the Planck constant and the pre-factor 2 accounts for spin degeneracy. The obtained values agree well with the densities measured in the same device using the Hall effect (inset of Fig. 3c). This indicates that all conduction electrons reside within a single electric sub-band and within a single valley, in agreement with the band structure calculations in Supplementary Section 7. This is in contrast to the behavior observed for 2DEGs formed at stacking faults in bulk InSe where, despite smaller carrier densities, four sets of SdHO were found originating from different electric sub-bands[25]. Only in 10L InSe, we observed some population of the second sub-band (Supplementary Section 4). The difference with bulk InSe is attributed to much stronger quantum confinement in our atomically thin crystals, which significantly increases the energy separation between sub-bands.

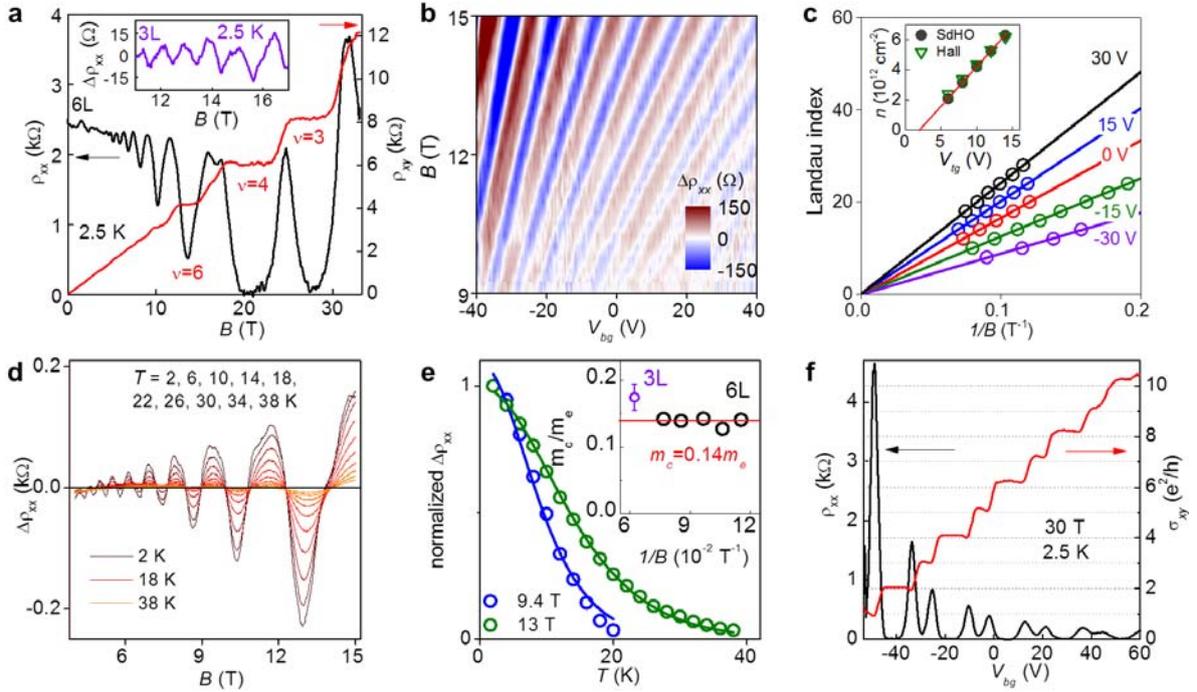

**Figure 3| Magnetotransport in few-layer InSe. a,** low-$T$ $\rho_{xx}$ and $\rho_{xy}$ for our 6L device; $n = 2 \times 10^{12}$ cm$^{-2}$. Inset: SdHO in 3L InSe; $n = 8.9 \times 10^{12}$ cm$^{-2}$ and smooth non-oscillating background is subtracted. **b,** Color map of SdHO amplitude, $\Delta\rho_{xx}(V_{bg}, B)$, in the 6L FED at $V_{tg} = 8$ V and 5 K. **c,** Minima in the SdHO, shown in **b**, as a function of $1/B$ for several $V_{bg}$. Inset: Carrier densities as a function of $V_{tg}$ determined from Hall measurements and SdHO. The red line is $n$ expected from the geometrical capacitance; $V_{bg} = -10$ V. **d,** Temperature dependence of $\Delta\rho_{xx}$ for 6L InSe; $n = 2.5 \times 10^{12}$ cm$^{-2}$. **e,** Examples of $\Delta\rho_{xx}(T)$, shown in **d**, at fixed $B$ and normalized by the corresponding values at 2 K. Solid lines: Best fits by the Lifshitz-Kosevich formula, which also yield the Dingle temperature of $\approx 10$ K. Inset: Cyclotron masses found for $N = 3$ and 6. Error bars for the 6L data are given by circles' size. The red line shows $m_c$ in bulk InSe. **f,** QHE effect in the 6L device as a function of $V_{bg}$ using $V_{tg} = 8$ V.

The SdHO were pronounced over a wide range of $T$ (Fig. 3d), which enabled us to determine the effective mass, $m_c$, of electrons in few-layer InSe. For 6L, the best fit of the SdHO amplitude using the Lifshitz-Kosevich formula yielded $m_c = 0.14 \pm 0.01\, m_e$ which is close to the bulk value[13] (Fig. 3e). A somewhat heavier $m_c \approx 0.17 \pm 0.02\, m_e$ was found for 3L InSe, where $m_e$ is the free electron mass. The high quality of the 2DEG in our 6L device allowed the observation of the fully developed QHE. Fig. 3f shows behavior of $\rho_{xx}$ and Hall conductivity $\sigma_{xy}$ with changing electric doping at $B = 30$ T. While $\sigma_{xy}$ exhibits plateaus close to the quantized values $\nu e^2/h$ (where ν is integer), the resistivity drops to zero for the corresponding intervals of $V_{bg}$. We observed quantum Hall plateaus for all odd and even $1 \leq \nu \leq 10$, indicating a lifted spin degeneracy. The temperature dependence of the minimum in $\rho_{xx}$ at $\nu = 3$ was carefully measured, which allowed a rough estimate for the effective $g$-factor as $g \approx 2$ (Supplementary Section 5).

Despite several attempts, we were unable to achieve sufficiently high conductance in the 1L devices to carry out successful four-probe measurements (Supplementary Section 2). Even at highest doping, our monolayer FEDs exhibited two-probe resistances larger than 1 MΩ. The field-effect mobility extracted in the two-terminal geometry was <0.1 cm²/Vs but note that, unlike the Hall mobility, this measure is likely to be dominated by changes in the contact resistance[24]. Except for graphene, all other monolayer crystals studied so far[2,28,30] exhibited similarly low electronic quality. This can be related to stronger degradation of the ultimately thin films[31]. However, other scenarios are also possible. For example, the large band gap in monolayer InSe (predicted theoretically and reported below) should favor a larger density of midgap states, making it difficult to fill them and reach the conduction band edge. In addition, larger gaps usually result in higher Schottky barriers and, therefore, a higher contact resistance can be expected for the graphene-InSe interface.

To gain further information about few-layer InSe, we employed photoluminescence (PL) spectroscopy. Fig. 4 shows the PL response found for hBN-encapsulated InSe crystals using laser excitation at photon energies of 2.3, 2.7 and 3.8 eV (see Methods). We studied 2D crystals with every $N$ from 1 to 8. For 2 to 8 layer InSe, their PL spectra showed two lines, $A$ at a lower and $B$ at a higher energy, whereas monolayer InSe exhibited only the high-energy peak. The inset of Fig. 4a plots the energy of the $A$ and $B$ lines for different $N$, with dots corresponding to the measured PL and open squares to the DFT calculations described in Supplementary Section 7. The progressive blue shift of the $A$ line with decreasing $N$ follows the trend reported previously[17,18]. To highlight the disappearance of the $A$ line in 1L InSe, Fig. 4b shows a PL intensity map for a device containing mono- and bi- layer regions. There was no discernable PL response in the $A$ spectral region anywhere within the 1L area.

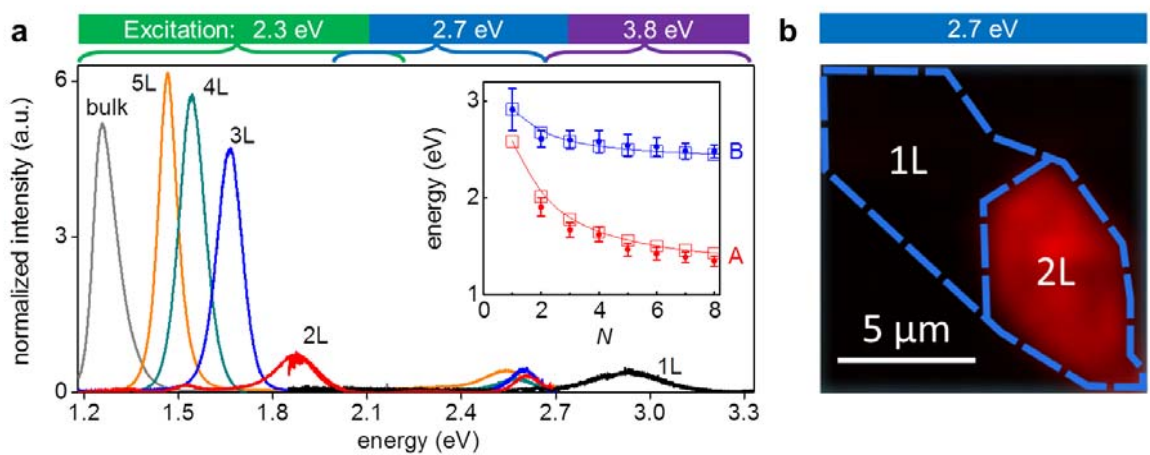

**Figure 4| Photoluminescence from 2D InSe. a,** PL spectra measured at RT using three excitation energies. The corresponding ranges of the PL detection are indicated by different colors at the top. Excitation at 2.3 eV generated the $A$ peak in $N$-layer crystals and bulk InSe ($\approx 45$ nm). Excitation by 2.7 and 3.8 eV lasers resulted in the $B$ peak for all $N$. The intensities are normalized by the number of layers, except for bulk InSe in which case the intensity is normalized to fit the scale. Note that the

spectral peak at 1.9 eV for 2L InSe was accessible in both 2.3 and 2.7 eV excitation measurements, and the two red curves overlap. Inset: Energies of the *A* and *B* peaks for different *N*, with the error bars indicating the PL linewidth. The squares, connected by solid curves, show our DFT results. **b**, Map of PL intensity integrated over the spectral range $1.85 - 2.7$ eV for excitation at 2.7 eV. The blue contours indicate the position of mono- and bi- layer regions. Black to red scale: Intensity variations from zero to maximum.

To explain the observed PL behavior, we have carried out DFT calculations of the band structure of 2D InSe using its monolayer as a building block for thicker films. The results are summarized in the inset of Fig. 4a and detailed in Supplementary Section 7. One can understand the observed decrease in the PL energies with increasing *N* as follows. The electronic bands of 1L InSe become split into *N* sub-bands in the *N*-layer crystal, and this reduces the energy difference between conduction and valence states in the vicinity of the Γ-point at which the band gap is smallest[18,19]. This leads to a decrease in the principal (optical) band gap that gives rise to the *A* peak. A smaller reduction occurs for the *B* peak that involves deeper, less affected valence bands (Supplementary Fig. 7). Furthermore, because of the mirror-plane symmetry ($z \rightarrow -z$) for 1L InSe, wavefunctions for electronic states near the edges of the valence and conduction bands are even and odd with respect to transformation $z \rightarrow -z$, respectively[19,23]. This makes the lowest-energy electron-hole transition (at ~2.6 eV as found by DFT; inset of Fig. 4a) optically inactive for in-plane polarized light[18,19] and only weakly coupled to *z*-polarized light. In few-layer InSe, the mirror-symmetry is broken (see Fig. 1a) which promotes coupling to in-plane polarized light and gives rise to the optical transitions responsible for the *A* line. Coupling to *z*-polarized light also increases with *N* (see Supplementary section 7). In contrast, two deeper valence bands at the Γ-point in 1L InSe have odd wavefunctions giving rise to the *B* peak at ~2.9 eV. The symmetry for this transition is largely unaffected by increasing *N*. The relatively high visibility of the *B* line in 2D InSe comes from the fact that, according to our DFT, wavefunctions of electronic states in different valence bands differ so significantly that electron-phonon relaxation between them and Auger recombination are greatly suppressed.

To conclude, hBN-encapsulation of atomically thin InSe in an oxygen- and moisture- free atmosphere allows high-quality optics and electron transport devices, which is difficult to achieve otherwise. Because InSe exhibits higher environmental stability than few-layer black phosphorous and higher RT mobility and lighter electron mass than few-layer dichalcogenides, our work indicates a promising playground for studying low-dimensional phenomena and an interesting venue for developing ultrathin-body high-mobility nanoelectronics. In terms of optics, monolayer InSe features strongly suppressed recombination of electron-hole pairs, which can be used to pump the system to high exciton densities potentially suitable for studying excitonic complexes and exciton condensation.

**Supplementary Information**

*Sample preparation* - InSe devices were prepared using mechanical exfoliation and hBN encapsulation, which were carried out in an inert atmosphere of a glovebox. Thin InSe crystallites were first exfoliated from Bridgman-grown bulk γ-InSe onto a 300 nm layer of polymethyl glutarimide (PMGI). Their thickness was identified in an optical microscope and selectively verified by AFM. Then we used the dry peel transfer technique[26,27] to pick up a chosen InSe crystal by a larger hBN flake attached to a polymer membrane. The resulting hBN-InSe stack was transferred onto a relatively thick (> 50 nm) hBN crystal prepared on top of an oxidized Si wafer. The latter hBN had two narrow ribbons of few-layer graphene prepared on top of it. During the final transfer, the crystals were aligned so that edges of the InSe overlapped with the FLG to provide an ohmic contact[5]. Such encapsulation not only protected ultrathin InSe during following fabrication but also cleansed its surfaces from contamination[28].

The assembled hBN/Gr/InSe/hBN heterostructure was removed from the glovebox Ar environment and patterned using electron beam lithography to create quasi-one dimensional contacts to the FLG[27]. As metallic contacts, we deposited 3 nm of Cr followed by 50 nm Au. The next round of e-beam lithography was used to define a top gate that was in a shape of a multiterminal Hall bar. The metal top gate then served as an etch mask for reactive ion etching, which translated the Hall bar shape into InSe. The final devices were fully protected from the environment, except for the exposed etched edges[26,27], and did not show any signs of deterioration over many months.

*PL measurements* - Several encapsulated InSe structures were prepared to analyze their PL response. Typically, they were multi-terraced flakes containing parts of different thickness from bulk InSe down to a monolayer. Photoluminescence measurements were performed at room temperature using three different optical setups in order to cover the widest possible spectral range. (1) To study PL between 1.2 and 2.3 eV we used HORIBA's Raman system XploRATMPLUS with a laser of wavelength 532 nm (spot size $\approx$ 1 μm, laser power of 1.35 mW, and the spectrometer grating of 600 groves/mm). (2) For the midrange energies from 1.8 to 2.7 eV, we used Renishaw system InVia equipped with a 457 nm laser (spot size $\approx$ 1 μm, power of 0.2 mW, 2,400 groves/mm). (3) To detect PL from monolayer InSe, it was necessary to extend the spectral range and we used Horiba LabRAM HR Evolution (UV laser at 325 nm with the beam spot size of $\approx$2 μm, power of 1.2 mW and grating of 600 groves/mm). This allowed us to detect photoluminescence in the range from 2.7 to 3.8 eV. For each setup, spectra were collected using the same acquisition parameters (time and focus distance), and an additional spectrum from an area near the tested sample (without InSe) was acquired as a reference. The reference signal was subtracted from the spectra from *N*-layer regions. The resulting curves were normalized per absorbing layer, as mentioned in the caption to Fig. 4. For 1L InSe, accurate normalization was not possible because of the absence of the lower-energy, *A* peak in PL. However, one can see that the intensity of the *B* peak in Fig. 4a does not vary significantly with *N* for other thicknesses, which makes the reported intensity for 1L correct, at least qualitatively.

## SUPPLEMENTARY MATERIAL

### S1. Graphene-InSe contacts

The gate-tunable work function of graphene enabled us to form ohmic contacts between few-layer graphene and 2D InSe[1]. Prior to measurements of the InSe's electronic properties, we characterized all contacts for each device using the three-terminal measurement geometry (see Fig. S1, inset). The contact resistance, $R_C$, typically varied between 5 and 30 kΩ·µm. Figure S1 shows examples of such measurements as a function of back gate voltage.

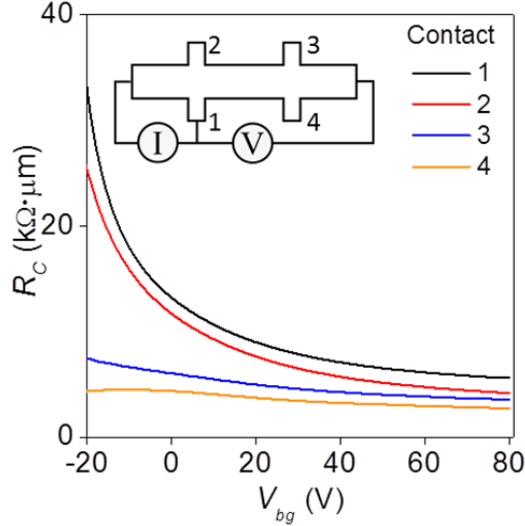

**Supplementary Figure 1| FLG contacts to 2D InSe.** The three-terminal resistance was measured using a contact in question (contact 1 for the case shown in the inset) as both current and voltage probe. A finite length of the InSe section next to the FLG interface can contribute to such measurements of $R_C$ but, using complementary four-probe measurements, we found that this contribution was relatively small and could be ignored. $V_{tg} = 5$ V, $I = 10$ nA and $T = 120$ K.

### S2. Electron transport in InSe monolayers

As mentioned in the main text, contacts to 1L InSe had high resistances, and we were able to carry out electrical measurements of the monolayers only in the two-probe geometry. Transport properties were studied by applying a source-drain voltage $V_{sd}$ and measuring the source-drain current $I_{sd}$ as a function of back gate voltage, $V_{bg}$. As shown in Fig. S2a, $I_{sd}$ grows with $V_{bg}$, yielding an on/off ratio of ~$10^2$ for this 1L device. In the off state, our measurements were limited by leakage currents. In the on state, the 2-terminal resistance decreased to only 1 MΩ at highest carrier concentrations we could reach. The field effect mobility, $\mu \approx 0.02$ cm²/Vs, was estimated using the relation $\mu = \left(\frac{L}{WCV_{sd}}\right)\frac{dI_{sd}}{dV_{bg}}$ where $L$ and $W$ are the length and width of the InSe channel, and $C$ is the capacitance per unit area of the back-gate dielectric (SiO$_2$ and hBN layers). Although the measured changes in resistance may be related to the field effect in monolayer InSe, we note that $R_C$ also varies with gate voltage and, therefore, can affect the two-probe measurement analysis. In principle, this contribution may be even dominant if the contact resistance is much larger than that of the InSe channel.

We also studied the photoresponse of monolayer InSe under illumination with a laser beam. Fig. S2b shows the $I_{sd}(V_{sd})$ dependence measured under illumination with different laser powers, $P$. The photocurrent $I_{ph} = I_{sd} - I_{sd-dark}$ follows a power law $I_{ph} \sim P^\gamma$ with $\gamma \sim 0.3$ (Fig. S2c). Figure S2d shows the observed time-dependent response for InSe monolayers, revealing a significant optical modulation of the current.

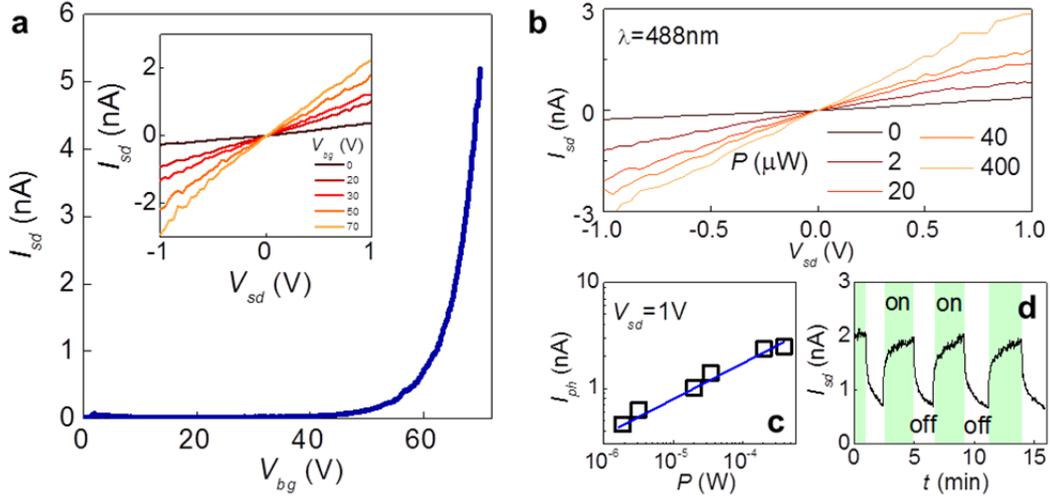

**Supplementary Figure 2| Characterization of monolayer InSe. a,** $I_{sd}$ as a function of gate voltage for one of our monolayer devices. Room $T$ and $V_{sd} = 5$ V. Inset: $I_{sd}(V_{sd})$ curves for different $V_{bg}$. **b,** $I_{sd}(V_{sd})$ dependence measured at $V_{bg} = 0$ V in the dark and under illumination with a 488 nm laser at different incident powers. **c,** Photocurrent as a function of laser power at $V_{sd} = 1$ V and $V_{bg} = 0$ V. The solid blue line represents the power law, $I_{ph} \sim P^{0.3}$. **d,** Time-resolved photo response under excitation at 633 nm ($V_{sd} = 1$ V and $V_{bg} = 0$ V).

## S3. Few-layer InSe in the two-probe geometry

Fig. 3a of the main text showed resistivity measurements for a 6L InSe device using the 4-probe configuration. In this case, the accessible range of gate voltages was limited by the contact resistance between graphene and InSe so that channel resistances of up to only a few kΩ could be examined. To demonstrate that much higher off-state resistances and, therefore, high on/off ratios could be achieved, we also studied this 6L device in the 2-probe geometry (Fig. S3). By varying $V_{tg}$ from 2 to 6 V, the two-terminal resistance changed from 4 GΩ to 40 kΩ, that is, the on/off ratio was $\sim 10^5$ with the off state being limited by electrical leakage. At all gate voltages the field-effect device was in the linear regime up to $V_{sd} \approx 15$ mV.

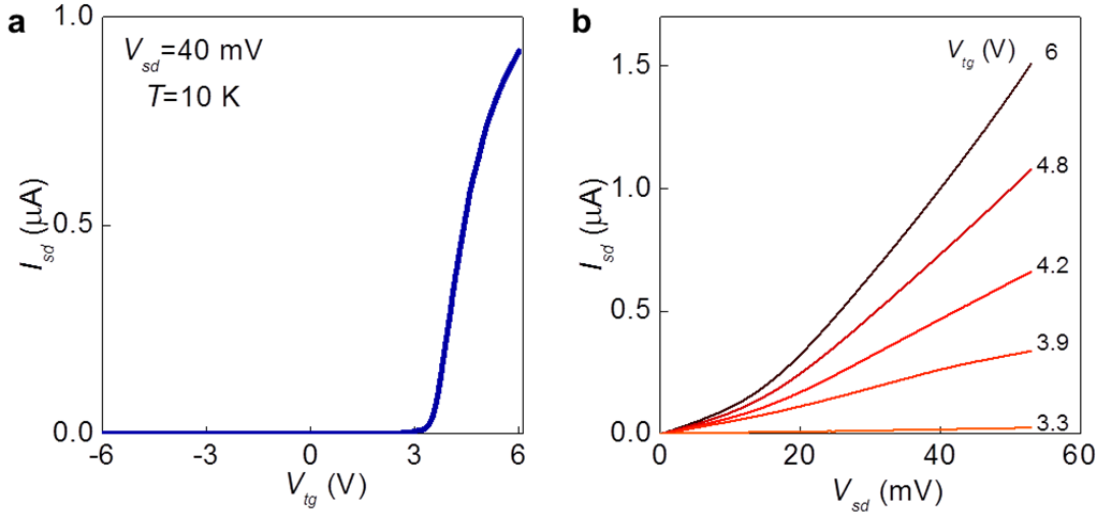

**Supplementary Figure 3| Field effect in 6L InSe using two-probe measurements. a,** $I_{sd}$ as a function of $V_{tg}$ ($V_{sd} = 40$ mV, $V_{bg} = -10$ V). **b,** $I_{sd}(V_{tg})$ at several $V_{tg}$ and measured for $V_{bg} = -10$ V and $T = 10$ K.

## S4. Further examples of electron transport in 2D InSe

Fig. S4 shows the resistivity of 3L and 10L InSe as a function of back gate voltage at different $T$. Independent of thickness, the resistivity grows with $T$ for all values of $V_{bg}$, similar to the case of 6L InSe discussed in the main text. The electron mobility μ for these devices was also sufficiently high to allow the observation of pronounced SdHO. Examples of $\rho_{xx}(B)$ and $\rho_{xy}(B)$ are shown in Fig. 4c where the oscillations start at around 6 T. One can notice that SdHO in this figure exhibit some additional beatings. Moreover, the carrier concentration determined from the Hall effect, $n = 7.8 \times 10^{12}$ cm$^{-2}$, differs from that estimated from the oscillations period, $n_1 = 5.65 \times 10^{12}$ cm$^{-2}$. To explain these observations, we carried out a Fourier transform analysis of $\rho_{xx}(B)$ and found that the beatings can be attributed to an additional oscillation frequency of $\approx 44$ T. It corresponds to the carrier density $n_2 = 2.12 \times 10^{12}$ cm$^{-2}$. The sum of two $n_1$ and $n_2$ equals the carrier density determined from the Hall effect. This behavior is attributed to the second electrical 2D subband that starts being occupied in 10L InSe, in contract to the single subband occupancy in 3L and 6L devices for the same carrier density.

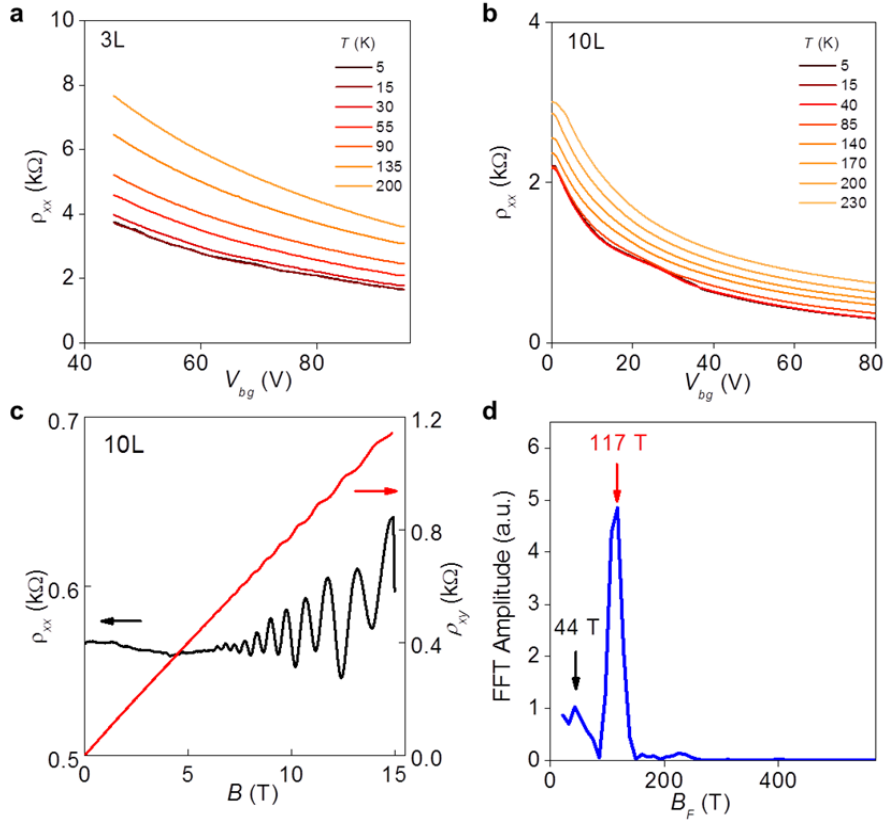

**Supplementary Figure 4 | Transport properties of few-layer InSe.** $\rho_{xx}$ as a function $V_{bg}$ for 3L (**a**) and 10L (**b**) InSe measured at $V_{tg} = 9.5$ V and $V_{tg} = 5$ V, respectively. **c,** $\rho_{xx}(B)$ and $\rho_{xy}(B)$ for 10L InSe ($V_{bg} = 80$ V and $V_{tg} = 5$ V; $T = 5$ K). **d,** Fourier transform of the SdHO shown in (**c**). Arrows indicate two fundamental frequencies that correspond to $n_1 = 5.65 \times 10^{12}$ cm$^{-2}$ and $n_2 = 2.12 \times 10^{12}$ cm$^{-2}$.

## S5. Estimating the effective g-factor

We have measured temperature dependence of $\rho_{xx}$ in the quantum Hall regime to get an estimate for the effective $g^*$-factor[2]. Its value is determined by the single-particle Zeeman energy and a many-body contribution arising from exchange splitting of electron spin states in partially occupied Landau level. Figure S5 shows $\rho_{xx}$ as a function of $V_{bg}$ at $B = 30$ T for our 6L InSe device. We observe a clear activation behavior $\rho_{xx}^{min} \sim \exp\left(-\frac{\Delta E}{k_B T}\right)$ of the minimum in $\rho_{xx}$ at $\nu = 3$ which yields $\Delta E/k_B \approx 9$ K where $k_B$ is the Boltzmann constant (Fig. S5 inset). Assuming no Landau level broadening, $\Delta E$ equals half of $g^*\mu_B B$ ($\mu_B$ is the Bohr magneton) and provides a lower bound for $g^*\mu_B B \approx 18$ K, so that $g^* \approx 0.9$. If we take into account Landau level broadening $\Gamma \approx \pi k_B T_D$, where $T_D = 9 \pm 2$ K is the Dingle temperature determined from SdHO (see Fig. 3c of the main text), the effective $g^*$-factor can be determined using relation $\Delta E = \frac{g^*\mu_B B - \Gamma}{2}$ which yields $g^*\mu_B B \approx 40$ K corresponding to $g^* \approx 2.3 \pm 0.3$. Further experiments in tilted magnetic field are required to improve the accuracy of determining the g-factor.

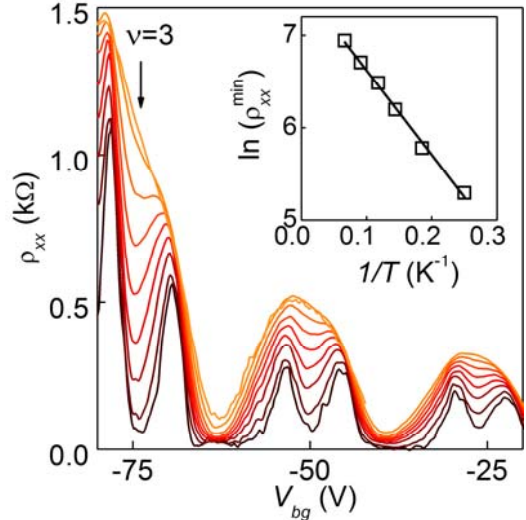

**Supplementary Figure 5| g-factor from temperature dependence of SdHO in 6L InSe.** $\rho_{xx}$ as a function of $V_{bg}$ for $T$ from 1.4 K (black) to 20.5 K (orange). $B$ = 30 T, $V_{tg}$ = 6.5 V. Inset: Log-scale of $\rho_{xx}^{min}$ versus $1/T$ at $\nu = 3$. The best linear fit (solid line) yields $\Delta E/k_B = 9$ K.

**S6. Stability of InSe flakes at ambient conditions**

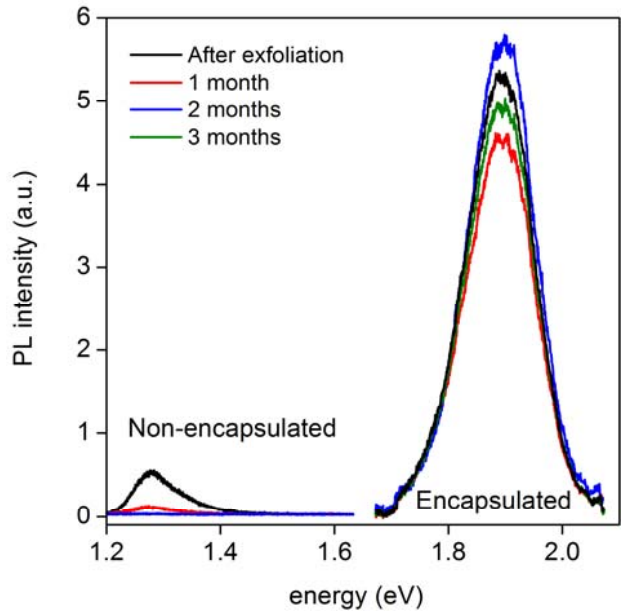

**Supplementary Figure 6| Stability of InSe thin films at ambient conditions.** PL response of encapsulated (2L) and non-encapsulated (22L) InSe films. Laser power 1.35 mW; excitation energy 2.33 eV.

Figure S6 shows photoluminescence (PL) from encapsulated (2L) and non-encapsulated (22L) InSe films measured immediately after preparation and a few months later. One can see that the PL response from the encapsulated sample remained practically unchanged, maybe even slightly improving with time due to annealing at room *T*. In contrast, the non-encapsulated InSe showed a significant drop of about 80% in the PL intensity already after one month of exposure to the air, despite being much thicker than the encapsulated bilayer. This translates into one layer of InSe being completely destroyed every couple of days, in agreement with the degradation rates reported in ref. 3. No PL signal could be detected from the non-encapsulated sample after two month under ambient conditions.

## S7. DFT modelling of the *N*-layer InSe band structure

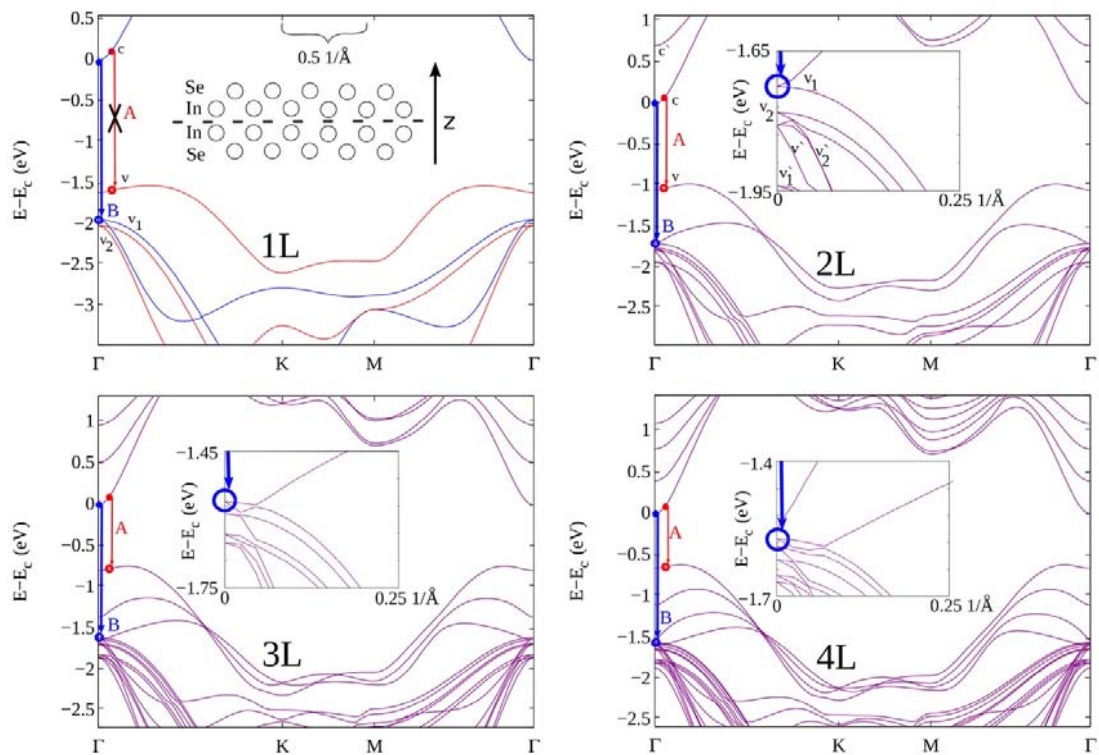

**Supplementary Figure 7| Band structure of InSe.** The electronic band structure of few-layer InSe, with zero energy set at the lowest (*c*) conduction band edge. In the monolayer (1L), the structure exhibits $z \to -z$ mirror symmetry, illustrated in the inset; we plot bands with $z \to -z$ symmetric wave functions in red and antisymmetric in blue. For two or more layers, this symmetry is no longer present. Principal optical transitions *A* and *B*, are marked by vertical lines, *A* in red and *B* in blue. For *N*L InSe (N=2,3,4) the insets expand the dispersion of overlapping bands in the vicinity of the Γ-point, with the valence subband involved in the *B*-transition identified by a circle.

Modelling of few-layer InSe was performed using semi-local density functional theory (DFT) with van der Waals corrections. We used the optB88 functional[4] as implemented in the VASP code[5] together with a plane-wave basis with a cut-off energy of 600 eV. We sampled the 2D Brillouin zone with a 12×12 grid. Note that the optB88 functional, being rooted in semi-local DFT, underestimates the band gap. Hence, when analysing the excitation energies, we apply the "scissor" correction, previously employed in studies of other semiconductors[6,7,8,9,10,11]. A calculation with optB88 gives a bulk InSe band gap $E_{g\infty}^{DFT} = 0.27$ eV, compared to the bulk experimental value of $\hbar\omega_{A\infty} = 1.25$ eV, measured at room temperature. Hence, to interpret the room temperature optics in the main text, we subtract

$$\delta E_{g\infty}^{300K} = \hbar\omega_{A\infty} - E_{g\infty}^{DFT} \approx 0.98 \text{ eV} \tag{1}$$

from the energies of all valence band states while keeping the conduction band energies unchanged. This is equivalent to adding $\delta E_{g\infty}^{300K}$ to the energies of all the interband transitions identified by analysing selection rules determined by the wave functions in the bands of monolayer and $N$L InSe. Note that the scissor correction should be temperature-dependent. For example, it has been found[12,13] that $\hbar\omega_{A\infty} = 1.4$ eV at low temperatures (T = 1.4 K), hence the scissor correction at low temperature should be 1.13 eV.

Fig. S7 shows the calculated band structure for 1L, 2L, 3L, and 4L InSe. For 1L InSe the lowest conduction band is labelled $c$, the highest valence band as $v$, the second and third highest valence band as $v_1$ and $v_2$, respectively. In 2L InSe each of these bands split into two subbands, which are traced by examining the wave functions of the bands at the Γ-point; higher energy conduction subbands and lower energy valence subbands are indicated by the symbols ($c`$, $v`$, $v_1`$, $v_2`$). The highest valence bands are "sombrero-shaped"[14,15,16]. The conduction band of monolayer InSe splits into several subbands, reducing the principal gap of $N$L InSe with increasing number of layers, while the electron effective mass changes only slightly. Two electron-hole excitations are highlighted in each panel, corresponding to the hole placed either at the top of the highest valence band which originates from the top valence band $v$ in 1L InSe (*A*-transition), or at the Γ-point in the highest of the bands generated by hybridization of the monolayer $v_1$ band (*B*-transition). We attribute these energies to the observed peaks in the photoluminescence (PL).

The PL intensity is governed by the matrix element for the optical transition (*A*- or *B*-type) and the occupation of photo-excited electron-hole states, both determined by the electron wave function in the bands. In monolayer InSe, the wave functions of any given band can be decomposed into orbitals

$$\psi_{\pm} = \frac{1}{\sqrt{2}} \sum_{orb} c_{orb,In}(\varphi_{orb,In1} \pm \varphi_{orb,In2}) + c_{orb,Se}(\varphi_{orb,Se1} \pm \varphi_{orb,Se2}), \tag{2}$$

where $c_{orb}$ are complex coefficients, $\varphi_{orb}$ are the atomic orbitals ($s$ or $p$) on the respective atoms, the set of orbitals in the decomposition are $orb=(s,p_x,p_y,p_z)$, and, in the case of $p_z$ orbitals, they are defined with opposite signs on the two atoms belonging to the same species in the unit cell: $\varphi_{pz,In2}(\vec{r} - \vec{R}_{In2}) = -\varphi_{pz,In1}(\vec{r} - \vec{R}_{In1})$. In this construction the wave functions exhibit mirror plane symmetry of the crystal: bands are either even ($\psi_+$) or odd ($\psi_-$) with respect to $z \to -z$ reflection. The optical transition for in-plane polarized light is determined by the matrix element ($\langle\psi_v|p_x|\psi_c\rangle$, $\langle\psi_v|p_y|\psi_c\rangle$) of the in-plane momentum operator, which is zero between even-odd bands.

|  | c (odd) | v (even) | $v_1$ (odd) | $v_2$ (even) |
|---|---|---|---|---|
| Degeneracy | 1 | 1 | 2 | 2 |
| $E^{DFT}$ (eV) | 0 | -1.60 | -1.93 | -2.01 |
| $E^{DFT} - \delta E_{g\infty}^{300K}$ (eV) |  | **-2.58** | **-2.91** | -2.99 |
| Se1 | 0.090$s$ | 0.002$s$ | 0.216$p_x$ | 0.207$p_x$ |
|  | 0.167$p_z$ | 0.349$p_z$ | 0.216$p_y$ | 0.207$p_y$ |
| In1 | 0.229$s$ | 0.028$s$ | 0.034$p_x$ | 0.043$p_x$ |
|  | 0.014$p_z$ | 0.121$p_z$ | 0.034$p_y$ | 0.043$p_y$ |
| In2 | 0.229$s$ | 0.028$s$ | 0.034$p_x$ | 0.043$p_x$ |
|  | 0.014$p_z$ | 0.121$p_z$ | 0.034$p_y$ | 0.043$p_y$ |
| Se2 | 0.090$s$ | 0.002$s$ | 0.216$p_x$ | 0.207$p_x$ |
|  | 0.167$p_z$ | 0.349$p_z$ | 0.216$p_y$ | 0.207$p_y$ |

**Supplementary Table S1| Orbital composition of wave functions in 1L-InSe.** Orbital decomposition of the wave functions at the Γ-point of conduction band (*c*), valence band (*v*), and doubly degenerate bands $v_1$ and $v_2$ (modulus square of the overlap integral between the DFT wave function and the spherical harmonics centered on each atom, $|c_{orb}|^2$ from Eq. 2) into *s* and *p* orbitals of In and Se atoms in monolayer InSe. The $z \to -z$ symmetry classification of the bands is noted in brackets. Atoms are listed from bottom to top for the 2L crystal. The Γ-point band energies are provided relative to the conduction band edge, where $E^{DFT}$ is the band energy from DFT and $E^{DFT} - \delta E_{g\infty}^{300K}$ is the value obtained after applying scissor correction $\delta E_{g\infty}^{300K}$. The energies corresponding to ℏω$_A$ and ℏω$_B$ are marked in bold.

The decomposition of the valence and conduction band states in 1L InSe to atomic orbitals, shown in Table S1, reveals that both states are dominated by *s* and $p_z$ orbitals, but in different combinations: *c* is dominated by *s* of In and $p_z$ of Se atoms, while *v* is dominated by $p_z$ orbitals of both In and Se atoms. Moreover, the valence band wave function is symmetric, and the conduction band wave function is antisymmetric to $z \to -z$ transformation; thus the optical matrix element vanishes. In *N*L InSe (e.g. bilayer, *N*=2), the stacking order is such that In atoms in the second layer are directly above Se atoms in the first layer, while Se atoms in the second layer are located above the centre of the hexagons in the first layer. This stacking order removes the $z \to -z$ mirror symmetry, which is reflected by the orbital decomposition of the band edge states. For example, in Table S2, the contribution of orbitals centred on atom In1 differs from that of orbitals centred on In2. This lower symmetry of few-layer InSe crystals allows the optical matrix element to be finite, which is reflected in Fig. S7 by marked *A*-transitions from the highest valence band to the lowest conduction band. We attribute this transition to the lowest energy PL line in NL InSe, discussed in the main text.

Also, in 1L InSe, there are two pairs of deeper valence bands, $v_1$ and $v_2$. Each pair is double-degenerate at the Γ-point and both are composed dominantly of $p_x$, and $p_y$ orbitals of Se, but $v_1$ is odd while $v_2$ is even with respect to $z \to -z$ reflection. Therefore, the mirror symmetry argument allows transition between *c* and $v_1$ bands, which we attribute to the *B* line observed in the PL experiments (see main text). In the bilayer, $v_1$ and $v_2$ are mixed, giving rise to four bands, all doubly degenerate at the Γ-point: $v_1$, $v_1$`, $v_2$, and $v_2$`. In $v_1$ and $v_1$`, the wave function is concentrated on the Se atoms on the inside of the bilayer, while in $v_2$, and $v_2$`, it is concentrated on the Se atoms on the outside. Among all these bands, we find (see below) that the strongest transition couples the band $v_1$ with the lowest conduction band, hence, we attribute this transition to the experimentally observed *B* line in PL. In Fig. S8, we gather the DFT-calculated values of the energies of *A*- and *B*-type electron-hole excitations, with the scissor-corrected values given by the energy scale on the right hand side axis.

| | c` | c | v | $v_1$ | $v_2$ | $v_2$` | v` | $v_1$` |
|---|---|---|---|---|---|---|---|---|
| Degeneracy | 1 | 1 | 1 | 2 | 2 | 2 | 1 | 2 |
| $E^{DFT}$ (eV) | 0.707 | 0 | -1.04 | -1.69 | -1.75 | -1.78 | -1.78 | -1.92 |
| $E^{DFT} - \delta E_{g\infty}^{300K}$ (eV) | | | **-2.02** | **-2.67** | -2.73 | -2.76 | -2.76 | -2.90 |
| Se1 | 0.051*s* 0.032$p_z$ | 0.043*s* 0.120$p_z$ | 0.006*s* 0.187$p_z$ | 0.054$p_x$ 0.054$p_y$ | 0.193$p_x$ 0.193$p_y$ | 0.167$p_x$ 0.167$p_y$ | 0.160$p_z$ | 0.009$p_x$ 0.009$p_y$ |
| In1 | 0.093*s* 0.051$p_z$ | 0.127*s* | 0.037*s* 0.050$p_z$ | 0.008$p_x$ 0.008$p_y$ | 0.034$p_x$ 0.034$p_y$ | 0.033$p_x$ 0.033$p_y$ | 0.002*s* 0.062$p_z$ | 0.003$p_x$ 0.003$p_y$ |
| In2 | 0.065*s* 0.003$p_z$ | 0.119*s* 0.018$p_z$ | 0.067$p_z$ | 0.029$p_x$ 0.029$p_y$ | 0.001$p_x$ 0.001$p_y$ | 0.012$p_x$ 0.012$p_y$ | 0.053*s* 0.052$p_z$ | 0.035$p_x$ 0.035$p_y$ |
| Se2 | 0.006*s* 0.174$p_z$ | 0.058*s* 0.035$p_z$ | 0.006*s* 0.131$p_z$ | 0.163$p_x$ 0.163$p_y$ | 0.011$p_x$ 0.011$p_y$ | 0.049$p_x$ 0.049$p_y$ | 0.018*s* 0.162$p_z$ | 0.201$p_x$ 0.201$p_y$ |
| Se3 | 0.006*s* | 0.055*s* | 0.008*s* | 0.162$p_x$ | 0.010$p_x$ | 0.050$p_x$ | 0.018*s* | 0.202$p_x$ |
|     | 0.179$p_z$ | 0.035$p_z$ | 0.131$p_z$ | 0.162$p_y$ | 0.010$p_y$ | 0.050$p_y$ | 0.162$p_z$ | 0.202$p_y$ |
| In3 | 0.065$s$ | 0.110$s$ | 0.075$p_z$ | 0.027$p_x$ | 0.001$p_x$ | 0.011$p_x$ | 0.053$s$ | 0.037$p_x$ |
|     | 0.006$p_z$ | 0.018$p_z$ |         | 0.027$p_y$ | 0.001$p_y$ | 0.011$p_y$ | 0.046$p_z$ | 0.037$p_y$ |
| In4 | 0.111$s$ | 0.114$s$ | 0.041$s$ | 0.007$p_x$ | 0.038$p_x$ | 0.030$p_x$ | 0.002$s$ | 0.003$p_x$ |
|     | 0.060$p_z$ |         | 0.054$p_z$ | 0.007$p_y$ | 0.038$p_y$ | 0.030$p_y$ | 0.059$p_z$ | 0.003$p_y$ |
| Se4 | 0.060$s$ | 0.038$s$ | 0.006$s$ | 0.051$p_x$ | 0.214$p_x$ | 0.148$p_x$ | 0.153$p_z$ | 0.009$p_x$ |
|     | 0.038$p_z$ | 0.109$p_z$ | 0.200$p_z$ | 0.051$p_y$ | 0.214$p_y$ | 0.148$p_y$ |         | 0.009$p_y$ |

**Supplementary Table 2| Orbital composition of wave functions in 2L-InSe.** Relative weights on the valence *s* and *p* orbitals of In and Se atoms in 2-layer InSe at the Γ-point ($|c_{orb}|^2$ from Eq. 2), for the bands labelled in Fig. S7. Atoms are listed from bottom to top for the 2L crystal. Band energies are provided relative to the lowest conduction band edge (*c*). $E^{DFT}$ is the Γ-point energy value obtained using DFT and $E^{DFT} - \delta E_{g\infty}^{300K}$ is the value obtained after subtracting the scissor correction $\delta E_{g\infty}^{300K}$ for room temperature. The energies corresponding to transitions ℏω$_A$ and ℏω$_B$ are marked in bold.

The interband transition coupling to light can be found using the DFT wave functions, by calculating the momentum matrix element,

$$\langle \psi_v | \vec{p} | \psi_c \rangle = \hbar \sum_i C_{i,v}^* C_{i,c} \vec{G}_i , \quad (3)$$

where c$_i$ are the plane-wave coefficients and $\vec{G}_i$ are the reciprocal lattice vectors evaluated using VASP. First, by calculating these matrix elements, we establish that all $c \to v_1$ transitions from the lowest conduction *c*-subband to the highest subband formed by interlayer coupling of monolayer *v$_1$*-bands are allowed in the in-plane polarization of light. The magnitude of the matrix element of such *B*-type transitions, $\beta = \sqrt{|\langle \psi_{v1}|p_x|\psi_c\rangle|^2 + |\langle \psi_{v1}|p_y|\psi_c\rangle|^2}$, is finite at the Γ-point and does not significantly depend on the number of layers. The calculated values of β are given in Table S3 for 1L, 2L, and 3L InSe.

The calculated optical matrix element of the *A*-transition (for up to 5 layers) is presented in Fig. S9, which shows that both $\langle \psi_v | p_x | \psi_c \rangle$ and $\langle \psi_v | p_y | \psi_c \rangle$ vanish at the Γ-point, as expected for non-degenerate bands in crystals with a hexagonal Bravais lattice. In the vicinity of the Γ-point, the matrix element is found to be linear in the electron wave vector, $\vec{k}$, in the band. This allows us to write down the Hamiltonian describing electrons near the edges of the band gap in *N*L InSe as

$$\hat{H}_N \approx \begin{pmatrix} \frac{(\hbar k)^2}{2m_{cN}} & \frac{\hbar \alpha_N e}{cm_e} \vec{k} \cdot \vec{A} + eE_z d_z \\ \frac{\hbar \alpha_N e}{cm_e} \vec{k} \cdot \vec{A} + eE_z d_z & -E_{gN} \end{pmatrix}, \quad (4)$$

where $\vec{A}$ and $E_z$ incorporate the coupling to the vector potential and out-of-plane electric field of a photon, respectively, $m_e$ is electron mass in vacuum, and $d_z = \langle \psi_v | z | \psi_c \rangle$. The *N*-dependence of the coefficient $\alpha$ and $d_z$ are shown in Table S3 and Fig. S9, respectively.

Finally, we note that the strong difference between wave functions in valence bands *v$_1$* and *v* leads to a slow interband relaxation, which is likely to be the reason for the experimentally observed hot luminescence in 1L, 2L, and 3L InSe; this requires the presence of long-living strongly non-equilibrium carriers in the optically pumped system.

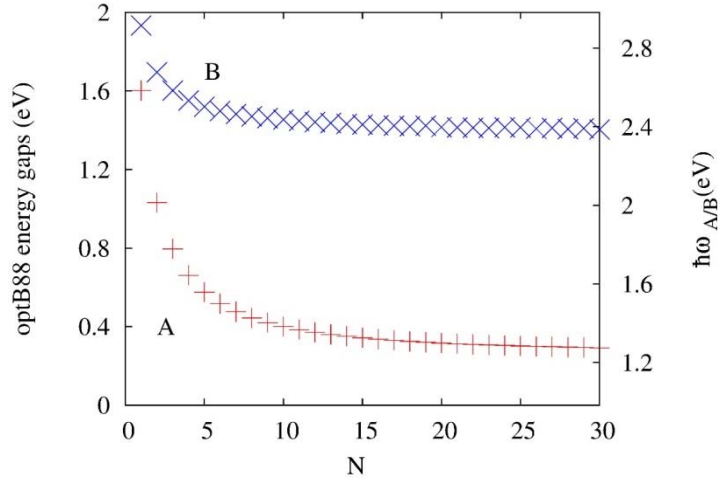

**Supplementary Figure 8|  Transition energies of InSe.** Energies of *A*- and *B*-transitions between the lowest conduction band *c* and the valence bands *v* and $v_1$, as marked in Fig. S7. The axis on the left shows the energy gaps obtained from DFT; the axis on the right shows the energies $\hbar\omega_{A/B}$ obtained after applying the scissor correction. The trend in the *A*-line energy dependence on the number of layers N agrees with the earlier theories [14,16,17].

| N | $E_{A/B}^{DFT}$ (eV) | $\hbar\omega_{A/B}$ (eV) | $m_c$ ($m_e$) | α | β ($\hbar$/Å) |
|---|---|---|---|---|---|
| 1 | 1.602 / 1.933 | 2.584 / 2.916 | 0.180 | 0.000 | 1.053 |
| 2 | 1.031 / 1.695 | 2.014 / 2.677 | 0.151 | 0.097 | 1.051 |
| 3 | 0.796 / 1.601 | 1.779 / 2.584 | 0.140 | 0.222 | 1.067 |

**Supplementary Table S3| *N*-layer InSe parameters:** DFT-calculated energy gap $E_{A/B}^{DFT}$ and transition energy $\hbar\omega_{A/B}$ obtained using scissor correction; the conduction band effective mass ($m_c$ in units of free electron mass); coefficients α and $\beta$ for the *A*- and *B*-transition.

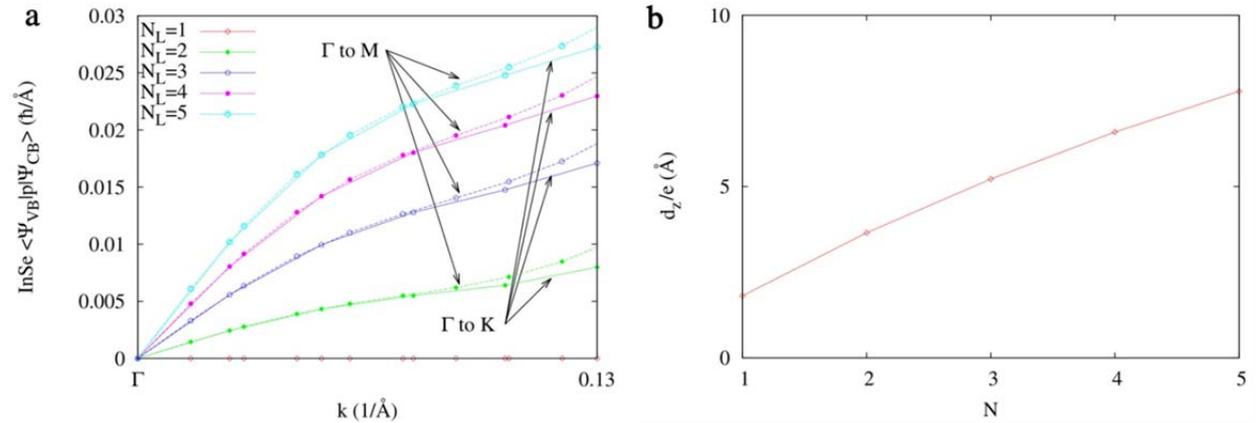

**Supplementary Figure 9| Optical matrix elements of InSe. a,** The optical transition matrix element for vertical transitions between the valence and conduction band in *N*L InSe, as a function of the modulus of the wave vector. **b,** The matrix element $d_z$ as a function of the number of layers $N$. Note that $d_z$ saturates at $N > 40$ to approximately $\frac{d_z}{e}(N = \infty) \approx 15$ Å .